\documentclass[10pt]{iopart}

\expandafter\let\csname equation*\endcsname\relax
\expandafter\let\csname endequation*\endcsname\relax 

\usepackage[english]{babel}
\usepackage{amsmath}
\usepackage{amssymb}
\usepackage{graphicx}
\usepackage{color}

\newcommand{\prob}{{\rm Prob}}
\newcommand{\probI}{P^{\rm I}}

\renewcommand{\ll}{{j}}
\newcommand{\zz}{{\ell}}
\newcommand{\n}{{n}}
\newcommand{\m}{{m}}

\begin{document}

\title{Universal statistics of longest lasting records of random walks and L\'evy flights}

\author{Claude Godr\`eche$^1$, Satya N. Majumdar$^2$, Gr\'egory Schehr$^2$}

\address{$^1$ Institut de Physique Th\'eorique, CEA Saclay and CNRS, 91191 Gif-sur-Yvette cedex, France\\
$\;^2$Laboratoire de Physique Th\'eorique et Mod\`eles Statistiques, UMR 8626, Universit\'e Paris Sud 11 and CNRS, B\^at. 100, Orsay F-91405, France}
\ead{claude.godreche@cea.fr,satya.majumdar@u-psud.fr,gregory.schehr@u-psud.fr}
\begin{abstract}
We study the record statistics of random walks after $\n$ steps, $x_0, x_1,\ldots, x_\n$, with arbitrary symmetric and continuous distribution $p(\eta)$
of the jumps $\eta_i = x_i - x_{i-1}$. 
We consider the age of the records, i.e. the time up to which a record survives. Depending on how the age of the current last record is defined, we propose three distinct sequences of ages (indexed by $\alpha$ = I, II, III) associated to a given sequence of records. 
We then focus on the longest lasting record, which is the longest element among this sequence of ages. 
To characterize the statistics of these longest lasting records, we compute: (i) the probability that the record of the longest age is broken at step $\n$, denoted by $Q^{\alpha}(\n)$, which we call the probability of record breaking and: (ii) the duration of the longest lasting record, $\ell_{\max}^{\alpha}(\n)$. 
We show that both $Q^{\alpha}(\n)$ and the full statistics of $\ell_{\max}^{\alpha}(\n)$ are universal, i.e. independent of the jump distribution $p(\eta)$. 
We compute exactly the large $\n$ asymptotic behaviors of $Q^{\alpha}(\n)$ as well as $\langle \ell_{\max}^{\alpha}(\n)\rangle$ (when it exists) and show that each case gives rise to a different universal constant associated to random walks (including L\'evy flights). 
While two of them appeared before in the excursion theory of Brownian motion, for which we provide here a simpler derivation, the third case gives rise to a non-trivial new constant $C^{\rm III} = 0.241749 \ldotsÉ$ associated to the records of random walks. Other observables characterizing the ages of the records, exhibiting an interesting universal behavior, are also discussed. 
\end{abstract}

\maketitle


\section{Introduction and main results}

\begin{figure}
\centering
\includegraphics[width=0.7\linewidth]{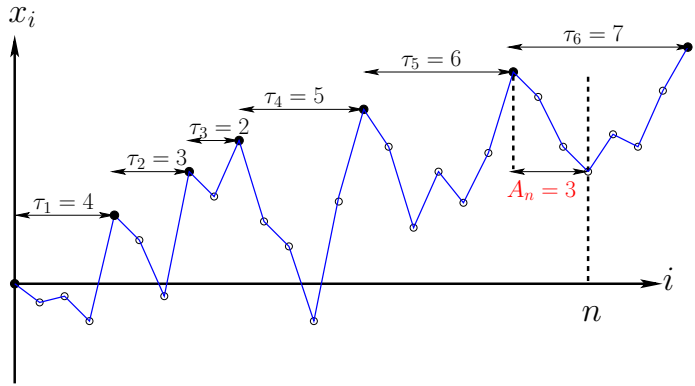}
\caption{One realization of a random walk of $\n=23$ steps, for which the number of records (the black dots) is $R_{23} = 6$.
The $\tau_k$ are the time intervals between successive records, or ages of the records,
and $A_\n$ is the age of the current record.
The next record occurs at step 27: $R_{27}=7$.}
\label{fig_record}
\end{figure}

The notion of records is becoming more and more popular in everyday life as, for instance, one hears and reads more and more often about ``record breaking events''. 
This is particularly true for sporting events where world or olympic records are often widely covered by the media \cite{Gembris}. 
In science, record statistics has found many applications in various areas, including in particular 
natural sciences and finance \cite{Nevzorov, ABN1992, DN2003} where extreme events might have drastic consequences. 
More recently, record statistics have been studied in statistical physics, where it was recognized that they also play an important role. Hence, there has
been a surge of interest for these questions in the physics literature. 
If one considers a discrete time series $x_0, \ldots, x_n$, where $x_i$'s might represent daily temperatures in a given city or the stock prices of a company, a record happens at time $k$ if the $k$-th entry is larger than all previous entries $x_0, \ldots, x_{k-1}$ (see Fig. \ref{fig_record}). 
One is naturally led to ask the following questions: (a) How many records occur in time $n$? (b) How long does a record survive and in particular what is the age of the longest lasting record? 
Such questions and related ones have found applications in various physical situations ranging   
from domain wall dynamics \cite{ABBM}, spin-glasses \cite{Sibani} and random walks \cite{MZ2008,satya_leuven,sanjib,WMS2012,MSW2012}, to avalanches \cite{LDW09}, models of stock prices \cite{WMS2012,WBK2011} or the study of global warming \cite{RP2006,WK2010}. They were also found to be relevant in evolutionary biology \cite{krugjain,franke} and in some random network growth process \cite{GL2008} (see Ref.~\cite{gregor_review, SM_review} for recent reviews). Recent investigations on the statistical mechanics of records revealed a rich phenomenology associated to rounding effects on record statistics \cite{rounding,edery} as well as first-passage behavior \cite{Redner_book, Satya_review,Bray_review} of record sequences \cite{BNK13,MBN13}.

Although the classical literature on records, chiefly from mathematical statistics \cite{Nevzorov, ABN1992, DN2003}, has mainly
focused on independent and identically distributed (i.i.d.) random variables, physical applications of records have instead emphasized the relevance of record statistics for strongly correlated variables, for which much less is known. Recently, it was realized that random walks (RW) is an ideal laboratory for analytical studies of record statistics \cite{MZ2008,satya_leuven,sanjib,WMS2012,MSW2012} (and more generally for extreme value statistics \cite{CM05,SM12,MMS13}) of strongly correlated variables. 
A RW generates a time series $x_0, \ldots, x_n$ via the following Markov process
\begin{eqnarray}\label{def_RW}
x_0 = 0 \;, \; x_i = x_{i-1} + \eta_i \;,
\end{eqnarray} 
where, in this work, $\eta_i$s are i.i.d. random variables distributed according to a symmetric probability density function (pdf) $p(\eta)$. Here we will restrict ourselves to the case where $p(\eta)$ is continuous, while the case of a discrete RW deserves a separate study \cite{MZ2008,SM_review}. 
Our study includes the cases where $p(\eta)$ has a narrow distribution (such that the RW converges to Brownian motion) as well as the case of L\'evy flights where $p(\eta)$ has fat tails, $p(\eta) \propto |\eta|^{-1-\mu}$, with $\mu < 2$ when $|\eta| \to \infty$. 

By definition, 
\begin{eqnarray}
x_k\; {\rm is \; a \; record \; if} \; x_k = \max{(x_1, \ldots, x_k)} \;,
\end{eqnarray}
and, by convention, we consider $x_0 = 0$ as the first record. 
A natural question concerns the number of records, $R_n \geq 1$, after $\n$ time steps. 
This question was first addressed in Ref.~\cite{MZ2008} where exact results were obtained for the full distribution of $R_\n$, 
\begin{equation}
P(\m,\n) = {\rm Prob}(R_\n = \m).
\end{equation}
Quite remarkably, it was shown that, thanks to the Sparre Andersen theorem \cite{SA53,SA54}, this distribution $P(\m,\n)$ is actually universal, i.e. does not depend on the jump distribution $p(\eta)$ for all values of $\n$ (hence, it holds also for L\'evy flights). In particular, for large $\n$, it was shown that $\langle R_\n \rangle \sim (2/\sqrt{\pi}) \sqrt{\n}$. These exact analytical results were later generalized to the case of $N$ independent random walkers \cite{WMS2012} and to the case of a single random walker with a drift \cite{MSW2012} where universality breaks down (at least partially) (see \cite{gregor_review,SM_review} for recent reviews). 

Apart from the number of records, other important observables which we focus on here are the ages of these records (see Fig. \ref{fig_record}). 
We thus define $\tau_k$ as the number of steps between the $k$-th and $(k+1)$-th records: this is the age of the $k$-th record, i.e. the time up to which the $k$-th record survives. 
Note that the last record still stays a record at step $\n$ and hence there are two distinct observables
associated with the duration of this last record: one is $\tau_{\m}$ --which is of course unknown at time $\n$-- and another one is the time $A_\n$ since the current record was set, or current age of the last record. 
Thanks to the translational invariance of the RW (see eq.~(\ref{def_ql}) and below), the sets of the ages $\tau_k$ together with $A_\n$ are in bijection with the intervals between two consecutive zeros of a RW. 
In this case, $\tau_k$ thus corresponds to the length of an excursion. 
Hence, as we will see below, the study of the ages of the records for a RW bears strong similarities with the excursion theory of RW and Brownian motion.

It is clear, as it is depicted in Fig. \ref{fig_record}, that the last record does not stand on an equal footing with the others (as there are for instance different ways to characterize its age). Here we are interested in studying the effects of this last record on various observables associated to the sequence of the ages. 
%
%
For this purpose, we propose here three distinct sequences which only differ by their last element, which allows to probe the sensitivity of the whole sequence on this last record. The first one, denoted as ``case I'' in the following, amounts to consider $A_\n$ as the relevant age of the last record. It is thus defined by the following sequence of ages ${\cal A}^{\rm I}_{\m,\n}$, where the indices $m,n$ in subscript refer to a number of records $R_n=m$ occurring in $n$ steps:
\begin{eqnarray}\label{def_I}
{\cal A}^{\rm I}_{\m,\n} = \{\tau_1, \tau_2, \ldots, \tau_{\m-1}, A_\n \} \;,
\end{eqnarray} 
while the second sequence, denoted as ``case II'' in the following, amounts to consider $\tau_\m$ as the relevant age of the last record. 
It is thus defined by
\begin{eqnarray}\label{def_II}
{\cal A}^{\rm II}_{\m,\n} = \{\tau_1, \tau_2, \ldots, \tau_{\m-1}, \tau_\m \} \;.
\end{eqnarray} 
Finally, to avoid the ambiguity of the age of the last record, one may simply discard it and consider instead a third ensemble, denoted as ``case III'' in the following:
\begin{eqnarray}\label{def_III}
{\cal A}^{\rm III}_{\m,\n} = \{\tau_1, \tau_2, \ldots, \tau_{\m-1}\} \;.
\end{eqnarray} 
The first ensemble (\ref{def_I}) is certainly the most natural one in the context of records, for instance, in sports and this ``case I'' was indeed considered in previous studies \cite{MZ2008,MSW2012}. 
The second ensemble (\ref{def_II}), as we shall see below, was studied in the excursion theory of random walks and Brownian motion \cite{Sch95}, using rather complicated probabilistic methods. 
\textcolor{black}{Finally, the last ensemble (\ref{def_III}) constitutes a toy model for the statistics of avalanches close to the depinning transition of elastic object in random media \cite{LDW09}. 
In this case, $\tau_k$ corresponds precisely to the size of the $k$-th avalanche for a system of size $\n$, while the quantity $A_\n$ in this context does not have a direct physical meaning. }
 
The typical fluctuations of these sequences (\ref{def_I}, \ref{def_II}, \ref{def_III}) is rather simple to study. For instance, the {\it typical} age of a record is simply given by $\ell_{\rm typ}(\n) \sim \n/\langle R_\n \rangle$. This {\it typical} behavior is thus the same for the three sequences (\ref{def_I}, \ref{def_II}, \ref{def_III}) as the probability distribution of $R_n$, ${\rm Prob}(R_n=m)$, remains unaffected by this choice $\alpha = {\rm I, II}$ and III. However, as we shall see, the statistics of the sequences of ages is instead dominated by {\it rare events}. 
It is the aim of the present paper to study the fluctuations of these rare events by addressing questions regarding the record ages of these three sequences. In this respect, an important quantity is the probability that the last time interval among the sequence is the longest one. 
This is the probability that the record of the longest age is broken at step $\n$. 
In the following, we will call this probability the ``probability of record breaking'' and we will denote it by $Q^{\rm \alpha}(\n)$, $\alpha={\rm I, \; II, \; III}$, depending on the three cases introduced above\footnote{$Q^{\alpha}(\n)$ should not be mistaken with the probability of record breaking of the time series itself (i.e. $Q^{\alpha}(\n) \neq {\rm Prob}\,[x_\n = \max(x_0, \ldots, x_\n)]$).}: 
\begin{eqnarray}
Q^{\rm \alpha}(\n) = 
\begin{cases}
&{\rm Prob} \, [A_\n \geq \max(\tau_1, \ldots, \tau_{\m-1})] \;, \; {\rm for \;} \alpha = {\rm I}\;,\\
&{\rm Prob} \, [\tau_\m \geq \max(\tau_1, \ldots, \tau_{\m-1})] \;, \; {\rm for \;} \alpha = {\rm II} \;,\\
&{\rm Prob} \, [\tau_{\m-1} \geq \max(\tau_1, \ldots, \tau_{\m-2})] \;, \; {\rm for \;} \alpha = {\rm III} \;.\\
\end{cases}\label{def_Q}
\end{eqnarray}
In case I (\ref{def_I}), this record breaking probability $Q^{\rm I}(n)$ was studied in the excursion theory of Brownian motion in \cite{PY97} where it was shown that (for a short review see~\cite{Fin2008})
\begin{equation}\label{Q_inf_I}
\lim_{\n \to \infty} Q^{\rm I}(\n) = Q^{\rm I}(\infty) = 
\int_{0}^\infty {\rm d}x\,\frac{1}{1+\sqrt{\pi x}\,\e^x {\rm erf} \sqrt{x} }
= 0.626508\ldots \;.
\end{equation}
Quite remarkably, it was recently shown that exactly the same constant $Q^{\rm I}(\infty)$ (\ref{Q_inf_I}) appears in the statistics of yet another extreme value observable associated to the sequence of ages (or equivalently of durations between successive zeros of a RW in the language of excursions) ${\cal A}^{\rm I}_{\m,\n}$ (\ref{def_I}). This observable is the age of the longest lasting record up to step $\n$, denoted by $\ell^{\rm I}_{\max}(\n)$
\begin{eqnarray}
\ell_{\rm max}^{\rm I} = \max(\tau_1, \ldots, \tau_{\m-1},A_\n) \;.
\end{eqnarray}
In Ref.~\cite{MZ2008} it was shown that the average longest age $\langle \ell_{\rm max}^{\rm I}(\n) \rangle$ grows linearly with $\n$, as in the case of i.i.d. random variables\footnote{In the case where $x_i$s are i.i.d. random variables, $\langle \ell_{\max}^{\rm I}(\n)\rangle \sim c_1 \n$ where $c_1 = 0.624329\ldots$ is the Golomb-Dickman constant \cite{FinBook}.}, but with an amplitude given in this case precisely by $Q_{\rm \infty}^{\rm I}$, i.e. \cite{MZ2008}
\begin{eqnarray}\label{lmax_I}
\lim_{\n \to \infty} \frac{\langle \ell_{\rm max}^{\rm I}(\n) \rangle}{\n} = C^{\rm I} = Q^{\rm I}(\infty) \;. 
\end{eqnarray}
The connection between $Q^{\rm I}(\n)$ and $\ell^{\rm I}_{\max}(\n)$ was eventually understood in Ref.~\cite{GMS2009} where it was shown that 
\begin{eqnarray}\label{relation_I}
\langle \ell^{\rm I}_{\max}(\n+1)\rangle = \langle \ell^{\rm I}_{\max}(\n)\rangle + Q^{\rm I}({\n}) \;,
\end{eqnarray}
which explains the occurrence of the same constant $Q^{\rm I}(\infty)$ for both quantities in (\ref{Q_inf_I}) and~(\ref{lmax_I}). 

Motivated by this connection between the probability of record breaking and the age of the longest lasting record in case I (\ref{relation_I}), we introduce the age of the longest lasting record $\ell_{\max}^{\rm \alpha}(\n)$, i.e. the largest element of the sequences ${\cal A}^{\rm \alpha}_{\m,\n}$ corresponding to the two other cases $\alpha$ = II, III:
\begin{eqnarray}
\ell_{\max}^{\rm \alpha}(\n) = 
\begin{cases}
&\max (\tau_1, \ldots, \tau_{\m-1}, \tau_\m) \;, \; {\rm for} \; \alpha = {\rm II}\;, \\
&\max (\tau_1, \ldots, \tau_{\m-1}) \;, \; {\rm for} \; \alpha = {\rm III} \;,
\end{cases}
\end{eqnarray} 
and a natural question is then whether $\langle \ell_{\max}^{\rm \alpha}(\n) \rangle$, for $\alpha$ = II, III, is related to $Q^{\rm \alpha}(\n)$ as it is in case $\alpha = $ I (\ref{relation_I}).

It is useful to summarize our main results. 
We show that the probability of record breaking $Q^{\rm \alpha}(\n)$, as well as the full statistics of $\ell_{\max}^{\alpha}(\n)$, are, firstly, different in each of the three cases considered here (\ref{def_I}, \ref{def_II}, \ref{def_III}) and, secondly, are universal in each case, i.e. independent of the jump distribution $p(\eta)$ of the $\eta_i$s in (\ref{def_RW}). The first result illustrates the striking fact that the statistics of the ages is extremely sensitive to its last element, the last record. On the other hand, the universal behavior of the full statistics of the ages, valid for any finite $\n$, stems from the universality of the Sparre Andersen theorem. 
Besides, although there is a simple relation between $Q^{\rm I}(\n)$ and $\langle \ell_{\max}^{\rm I}(\n) \rangle$ in case I (\ref{relation_I}), 
we show that such a relation does not exist between $\langle \ell_{\max}^\alpha(\n)\rangle$ and $Q^{\alpha}(\n)$ in the two other cases $\alpha =$ II, III. Finally, in the limit $\n \to \infty$, we find that these quantities are characterized by universal quantities, including in particular three universal constants which are summarized in Table \ref{Table_1}. 
\begin{table}
\begin{center}
\begin{tabular}{|c||c|c|}
\hline
Case & $Q^{\alpha}(\n)$ & $\langle \ell_{\max}^{\alpha}(\n)\rangle/\n$ \\
\hline
$\quad$ & $\quad$ & $\quad$ \\
$\alpha$ = I & $\sim Q^{\rm I}(\infty) = 0.626508\ldots$ & $\sim C^{\rm I} = Q^{\rm I}(\infty)$ \\
$\quad$ & $\quad$ & $\quad$ \\
\hline
$\quad$ & $\quad$ & $\quad$ \\
$\alpha$ = II & $\sim Q^{\rm II}(\infty) = 0.800310\ldots$ & $\infty$ \\
$\quad$ & $\quad$ & $\quad$ \\
\hline
$\quad$ & $\quad$ & $\quad$ \\
$\alpha$ = III & $\sim  \dfrac{\ln \n+c}{2\sqrt{\pi\n}}$ & $\sim C^{\rm III} = 0.241749\ldots$ \\
 $\quad$ & $\quad$ & $\quad$ \\
\hline
\end{tabular}
\caption{Asymptotic behaviors, when $\n \to \infty$, of the two main quantities studied here: the record breaking probability $Q^{\rm \alpha}(\n)$ and the average age of the longest lasting record up to time $\n$, $\langle \ell_{\max}^{\rm \alpha}(\n) \rangle$ in each three cases (\ref{def_I}, \ref{def_II}, \ref{def_III}).
The constant $c$ is defined in~(\ref{Q3_asympt}).}
\label{Table_1}
\end{center}
\end{table}
Case I was discussed previously in Ref.~\cite{MZ2008,GMS2009,PY97} and we have already seen that it is characterized by a single universal constant $Q^{\rm I}(\infty)$ (\ref{Q_inf_I}). In the second case, we show that the record breaking probability $Q^{\rm II}(\n)$ tends asymptotically to a finite constant, 
\begin{equation}\label{eq:scheffer}
\lim_{\n \to \infty} Q^{\rm II}(\n)=Q^{\rm II}(\infty) = \frac{1}{2} \int_0^\infty {\rm d}x\, \frac{\e^x-1}{x + \sqrt{\pi }x^{3/2} \, \e^x \, {\rm erf}(\sqrt{x})} = 0.800310 \ldots \;,
\end{equation}
where ${\rm erf}(x) = (2/\sqrt{\pi}) \int_0^x {\rm d}t\, \e^{-t^2} $ denotes the error function.
This constant appeared before in Ref.~\cite{Sch95} in the context of the record duration of the excursions of Brownian motion. 
There, this constant $Q^{\rm II}(\infty)$ was computed using rather complicated probabilistic methods. 
Here, we present a much simpler derivation of this result and generalize it to any RW with symmetric and continuous jump distribution. Finally, in the third case we show that $Q^{\rm III}(\n)$ does not converge to a strictly positive constant when $\n \to \infty$ but decays instead to $0$ in a non trivial and universal manner, 
\begin{equation}
Q^{\rm III}(\n) \sim \frac{\ln \n+c}{2{\sqrt{\pi\n}}},
\end{equation}
(see Table \ref{Table_1}), where the constant $c$, defined below in~(\ref{Q3_asympt}), is universal.
In this third case, the statistics of $\ell_{\max}^{\rm III}(\n)$ gives rise to a third universal constant associated to the record statistics of RW, $C^{\rm III}$, given by
\begin{equation}\label{C_III}
\lim_{\n \to \infty} \frac{\langle \ell_{\max}^{\rm III}(\n) \rangle}{\n} = C^{\rm III} \;, \; 
C^{\rm III} = \int_0^\infty {\rm d}x\,\frac{1-\sqrt{\pi x}\, \e^x\, {\rm erfc}(\sqrt{x})}{1+\sqrt{\pi x} \, \e^x \, {\rm erf}(\sqrt{x})}  = 0.241749 \ldots \;,
\end{equation} 
a constant which we have not seen before in the literature on RW and Brownian motion. 
In eq.~(\ref{C_III}), ${\rm erfc}(x) $ denotes the complementary error function ${\rm erfc}(x) = 1 - {\rm erf(x)}$. 

Finally we show that the two salient features of the statistics of the ages, namely its extreme sensitivity to the last record and its universality, are not restricted to $\ell^{\alpha}_{\rm max}(n)$ and $Q^{\alpha}(n)$ but also hold for other observables: this includes the shortest age $\ell_{\rm min}^{\alpha}(n)$ --i.e. the smallest element among the sequences of ages ${\cal A}^\alpha_{m,n}$-- and the probability $Q_1^{\alpha}(n)$ that the longest age is the first one, in contrast to the last one as in the definition of $Q^\alpha(n)$ in (\ref{def_Q}). The asymptotic behaviors of these observables are summarized in Table \ref{Table_2}.

\section{General framework}

The probability of record breaking $Q^{\alpha}(n)$ as well as the statistics of $\ell_{\rm \max}^\alpha(n)$ are both obtained from the joint distribution of the ages $\tau_k$ and $A_n$ (the precise ensemble of ages depends on the sequences ${\cal A}^{\alpha}_{m,n}$, $\alpha$ = I, II, III) and the number of records $R_n =m$. 
This joint distribution is generically denoted by 
\begin{equation}
P^{\alpha}(\vec \ell, m,n)=\prob(\vec\tau=\vec\ell,R_n=m) \;,
\end{equation}
with $\alpha = {\rm I, II, III}$, and where $\vec \zz$ is a realization of the set of ages $\vec\tau$. 
To compute $P^{\alpha}(\vec \ell, m,n)$ we need two quantities as input~\cite{MZ2008}. 
The first one is the probability $q(\ll)$ that a RW, starting at $x_0$, stays below $x_0$ after $\ll$ time steps:
\begin{eqnarray}\label{def_ql}
q(\ll) = {\rm Prob} (x_k < x_0, \; \forall \; 1 \leq k \leq \ll) \;,
\end{eqnarray}  
and we define $q(0)=1$.
Due to translational invariance, this probability is independent of $x_0$ and we can thus set $x_0=0$. 
Its generating function (GF) is given by the Sparre Andersen theorem~\cite{SA53}:
\begin{eqnarray}\label{SA_th}
\tilde q(z) = \sum_{\ll\ge0} q(\ll) z^\ll = \frac{1}{\sqrt{1-z}} \Longrightarrow q(\ll) = {2\ll \choose \ll} \frac{1}{2^{2\ll}} \;.
\end{eqnarray}
Thus
\begin{equation}
q(0)=1,\;q(1)=\frac{1}{2},\;q(2)=\frac{3}{8},\;q(3)=\frac{5}{16}, \ldots
\end{equation}

The second quantity which we will need is the first passage probability $f(\ll)$ that the RW crosses its starting point $x_0$ between steps $(\ll-1)$ and $\ll$ from below $x_0$. 
It follows from its definition that $f(\ll) = q(\ll-1) - q(\ll)$ so that its GF can be expressed~as
\begin{eqnarray}
\tilde f(z) = \sum_{\ll\ge1} f(\ll) z^\ll = 1 - (1-z)\tilde q(z) = 1 - \sqrt{1-z}\;.
\end{eqnarray} 
Thus
\begin{equation}
f(1)=\frac{1}{2},\;f(2)=\frac{1}{8},\;f(3)=\frac{1}{16},\ldots
\end{equation}
Case I was already treated in Ref.~\cite{MZ2008,GMS2009,PY97} and though the focus here is on cases II and~III we recall what is known on case I and add a few complements. 

\bigskip{\bf Case I}. 
In this case, a realization of the set of the ages is $\vec \ell = (\ell_1, \ell_2, \ldots, \ell_{m-1},a)$,
and the joint distribution of the ages and of $R_n$ is given by
\begin{equation}\label{joint_caseI}
\probI(\vec \ell, m,n)=f(\ell_1) f(\ell_2) \ldots f(\ell_{m-1}) q(a) \,
\delta\Big(\sum_{k=1}^{m-1}\ell_k+a,n\Big) \;,
\end{equation}
where $\delta(j,k)$ denotes the Kronecker delta. Summing this distribution on $\ell_1, \ell_2, \ldots, \ell_{m-1}$, from 1 to $\infty$, and upon $a$ from 0 to $\infty$
yields $\prob(R_n=m)$, whose GF reads~\cite{MZ2008}
\begin{equation}
\sum_{n\ge0}\prob(R_n=m)z^n=\tilde f(z)^{m-1}\,\tilde q(z),
\end{equation}
with $m=1,\ldots,n+1$.

The cumulative distribution function of $\ell_{\max}^{\rm I}(n)$ reads
\begin{eqnarray}\label{def_FI}
&&F^{\rm I}(\zz,n) = {\rm Prob} \, (\ell_{\max}^{\rm I}(n) \leq \zz) = \sum_{m\ge1} F^{\rm I}(\zz,m,n) \;,
\end{eqnarray}
where
\begin{eqnarray}
F^{\rm I}(\zz,m,n) 
&&= \prob \, (\ell_{\max}^{\rm I}(n) \leq \zz,R_n=m)\\
&&=\sum_{\ell_1=1}^\zz \cdots \sum_{\ell_{m-1}=1}^\zz \sum_{a=0}^\zz P^{\rm I}(\vec \ell, m,n) \;.
\end{eqnarray}
The GF with respect to $n$ of this last quantity is thus given by 
\begin{eqnarray}
\sum_{n\ge0}  F^{\rm I}(\zz,m,n) z^n= 
\Bigg(\sum_{\ll=1}^\zz f(\ll) z^\ll\Bigg)^{m-1}\sum_{j=0}^{\zz}q(j)z^j \;,
\end{eqnarray}
and therefore, summing on $m$,
\begin{eqnarray}
\tilde F^{\rm I}(z,\zz)=\sum_{n\ge0}  F^{\rm I}(\zz,n) z^n=  \frac{\sum_{j=0}^{\zz}q(j)z^j}{1 - \sum_{\ll=1}^\zz  f(\ll)z^\ll} \;.
\end{eqnarray}
The normalization can be checked in these two last expressions by letting $\zz\to\infty$, reminding that $\tilde f(z) = 1 - (1-z)\tilde q(z)$.

From the cumulative distribution of $\ell_{\max}$, $F^{\rm I}(\zz,n)$, we obtain its mean value $\langle\zz^{\rm I}_{\max}(n)\rangle$ by summation over $\ell$:
\begin{equation}
\langle\zz^{\rm I}_{\max}(n)\rangle=\sum_{\zz\ge0}(1-F^{\rm I}(\zz,n)),
\end{equation}
hence its GF is given by
\begin{eqnarray}\label{eq:lmaxI}
\sum_{n\ge0}\langle\zz^{\rm I}_{\max}(n)\rangle z^n
&&=\sum_{\zz\ge0}\Big(\frac{1}{1-z}-\tilde F^{\rm I}(z,\zz)\Big),\\
&&=z+\frac{3}{2}z^{2}+\frac{17}{8}z^{3}+\frac{11}{4}z^{4}+\cdots.\label{eq:coeff_lmax}
\end{eqnarray}
As we will see later on, $\langle\zz^{\rm I}_{\max}(n)\rangle$ scales as $n$ for large $n$.

Similarly, the probability of record breaking $Q^{\rm I}(n)$ can be obtained from the joint distribution $P^{\rm I}(\vec \ell, m,n)$ by summing over the number of records of the random walk,~i.e. 
\begin{equation}\label{eq:QI}
Q^{\rm I}(n) = \sum_{m\ge1} Q^{\rm I}(m,n) \;, \\
\end{equation}
where
\begin{eqnarray}
Q^{\rm I}(m,n) &&= \prob(A_n\ge\max(\tau_1,\ldots,\tau_{m-1}),R_n=m),\\
&&= \sum_{a\ge0} \sum_{\ell_1=1}^{a} \ldots 
\sum_{\ell_{m-1}=1}^{a} P^{\rm I}(\vec \ell, m,n) \;.
\end{eqnarray}
Its GF with respect to $n$ reads
\begin{eqnarray}\label{Q1_gf}
\tilde Q^{\rm I}(z)=
\sum_{n\ge0}  Q^{\rm I}(n) z^n=  \sum_{j\ge0} \frac{q(j)z^j}{1 - \sum_{k=1}^j f(k) z^k} \;.
\end{eqnarray}
The first orders $Q^{\rm I}(1), Q^{\rm I}(2), \ldots$ can be read off from the first terms of this series
\begin{equation}\label{eq:coeffQ}
\tilde Q^{\rm I}(z)=
1 +\frac{1}{2}z +\frac{5}{8}z^{2} +\frac{5}{8}z^{3} +\frac{81}{128}z^{4} +\frac{5}{8}z^{5} +\frac{161}{256}z^{6} 
+\cdots
\end{equation}
The coefficients of this series converge to a constant, as depicted in~figure~\ref{fig_QI},
which is $Q^{\rm I}(\infty)=0.626508\ldots$ given in~(\ref{Q_inf_I}) (see below). 
Using the identity
\begin{eqnarray}\label{eq:denominator}
1 - \sum_{\ll=1}^\zz f(\ll)z^\ll = q(\zz) z^\zz+ (1-z)\sum_{\ll=0}^{\zz-1}q(\ll)z^\ll  \;,
\end{eqnarray}
obtained from the definition $f(\ll) = q(\ll-1)-q(\ll)$ with $q(0)=1$,
we observe that
\begin{eqnarray}
(1-z) \sum_{n\ge0} \langle \ell^{\rm I}_{{\rm \max}}(n)\rangle  z^n= z \sum_{n\ge0} Q^{\rm I}(n)z^n \;,
\end{eqnarray}
which is in agreement with~(\ref{relation_I}). 
This latter result can itself be checked by comparing the coefficients of the two series~(\ref{eq:coeff_lmax}) and~(\ref{eq:coeffQ}).

\begin{figure}
\centering
\includegraphics[width=0.7\linewidth]{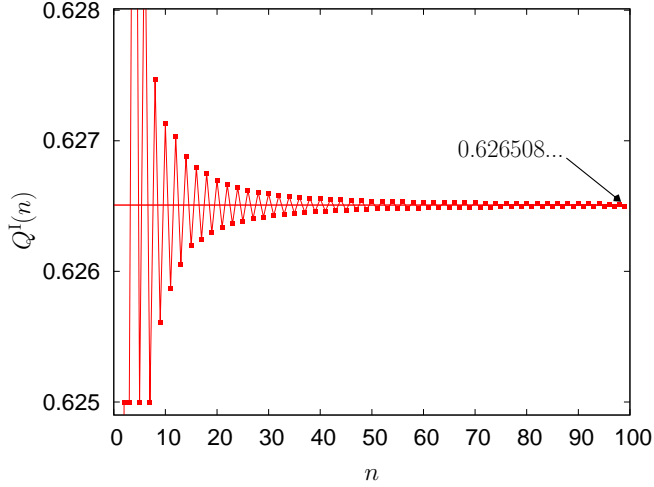}
\caption{The successive coefficients $Q^{\rm I}(n)$ of the series~(\ref{Q1_gf}) converge to $Q^{\rm I}(\infty)=0.626508\ldots$.
Odd and even coefficients are intertwined. }
\label{fig_QI}
\end{figure} 

%
\bigskip{\bf Case II}. 
In this case, a realization of the relevant set of ages is $\vec \ell = (\ell_1, \ell_2, \ldots, \ell_m)$ and the joint probability $P^{\rm II}(\vec \ell, m,n)$ is thus given by
\begin{equation}\label{eq:joint_pdf_II}
P^{\rm II}(\vec \ell, m,n) = f(\ell_1) f(\ell_2) \ldots f(\ell_{m-1}) f(\ell_m) 
I\left(\sum_{k=1}^{m-1} \ell_k\le n<\sum_{k=1}^m \ell_k \right) \;,
\end{equation}
where $I(\cdot)=1$ if the condition inside the parentheses is satisfied, and $I(\cdot)=0$ otherwise
%
and where we have used the Markov property of the RW. This formula (\ref{eq:joint_pdf_II}) implies that the intervals $\tau_k$ are statistically independent except for an overall global constraint that there are exactly $m$ records up to time $n$ and that $\ell_m$ is thus the duration between the $m$-th and $(m+1)$-th records. 
This constraint is ensured in (\ref{eq:joint_pdf_II}) by the indicator function $I(\cdot)$. 

As for case I, we can recover the generating series of $\prob(R_n=m)$ with respect to $n$
by summing on $\ell_1, \ell_2, \ldots, \ell_{m}$, from 1 to $\infty$ in~(\ref{eq:joint_pdf_II}), thus obtaining
\begin{equation}
\sum_{n\ge0}\prob(R_n=m)z^n
=\tilde f(z)^{m-1}\sum_{\ell_m\ge1}f(\ell_m)\frac{1-z^{\ell_m}}{1-z}
=\tilde f(z)^{m-1}\,\tilde q(z).
\end{equation}
The cumulative distribution function of $\ell_{\max}^{\rm II}(n)$ reads
\begin{equation}\label{def_FII}
F^{\rm II}(\zz,n) = {\rm Prob} \, (\ell_{\max}^{\rm II}(n) \leq \zz) = \sum_{m\ge1} F^{\rm II}(\zz,m,n) \;,
\end{equation}
where
\begin{eqnarray}
F^{\rm II}(\zz,m,n) 
&&= \prob \, (\ell_{\max}^{\rm II}(n) \leq \zz,R_n=m)\\
&&=\sum_{\ell_1=1}^\zz \cdots \sum_{\ell_{m-1}=1}^\zz \sum_{\ell_m=1}^\zz P^{\rm II}(\vec \ell, m,n) \;.
\end{eqnarray}
The GF of this last quantity with respect to $n$ is given by 
\begin{eqnarray}
\sum_{n\ge0}  F^{\rm II}(\zz,m,n) z^n= 
\Bigg(\sum_{\ll=1}^\zz f(\ll) z^\ll\Bigg)^{m-1}\sum_{j=1}^{\zz}f(j)\frac{1-z^j}{1-z} \;,
\end{eqnarray}
and therefore, summing on $m$,
\begin{eqnarray}\label{eq:FII}
\tilde F^{\rm II}(z,\zz)=\sum_{n\ge0}  F^{\rm II}(\zz,n) z^n
= \frac{1}{1-z} \frac{\sum_{\ll=1}^\zz f(\ll)(1 - z^\ll)}{1 - \sum_{\ll=1}^\zz  f(\ll)z^\ll} \;.
\end{eqnarray}
The normalization can be checked in these two last expressions by letting $\zz\to\infty$.

In this case, it turns out that the first moment $\langle \ell_{\rm \max}^{\rm II}(n) \rangle$ is not defined.
This is due to the fact that the pdf of $\ell_{\max}^{\rm II}(n)$ has a heavy tail. 
Indeed, the last excursion of the sequence~(\ref{def_II}), $\tau_m$, can be arbitrarily large. 
Hence the configurations with a large $\ell_{\max}^{\rm II}(n)$ are such that $\ell_{\max}^{\rm II}(n) = \tau_m$. 
Given that the pdf of $\tau_m$ has a power-law tail $\propto \tau_m^{-3/2}$, the pdf of $\ell_{\max}^{\rm II}(n)$ therefore inherits this tail with exponent $3/2$, implying that the first moment $\langle \ell_{\rm \max}^{\rm II}(n) \rangle$ is divergent (see Table \ref{Table_1}). 

The probability of record breaking $Q^{\rm II}(n)$ can be obtained from the joint distribution $P^{\rm II}(\vec \ell, m,n)$~(\ref{eq:joint_pdf_II}),
 by summing over the number of records of the random walk, i.e. 
\begin{equation}\label{Q2}
Q^{\rm II}(n) = \sum_{m\ge1} Q^{\rm II}(m,n) \;, \\
\end{equation}
where
\begin{eqnarray}
Q^{\rm II}(m,n) &&= \prob(\tau_m\ge\max(\tau_1,\ldots,\tau_{m-1}),R_n=m),\\
 &&= \sum_{\ell_m \ge1} \sum_{\ell_1=1}^{\ell_m} \ldots 
\sum_{\ell_{m-1}=1}^{\ell_m} P^{\rm II}(\vec \ell, m,n) \;.
\end{eqnarray}
Its GF with respect to $n$ reads
\begin{eqnarray}\label{Q2_gf}
\tilde Q^{\rm II}(z)=
\sum_{n\ge0} Q^{\rm II}(n) z^n = \frac{1}{1-z} \sum_{j\ge1} \frac{f(j)(1 - z^j)}{1 - \sum_{k=1}^j f(k) z^k} \;,
\\
=
1 +z+\frac{15}{16}z^2  +\frac{115}{128}z^{3} +\frac{1785}{2048}z^{4} +\frac{28053}{32768}z^{5} 
+\cdots\;.
\end{eqnarray}
The coefficients of this series converge to a constant, as depicted in~figure~\ref{fig_QII},
which is $Q^{\rm II}(\infty)=0.800310\ldots$ given in~(\ref{eq:scheffer}) (see below).

\begin{figure}
\centering
\includegraphics[width=0.7\linewidth]{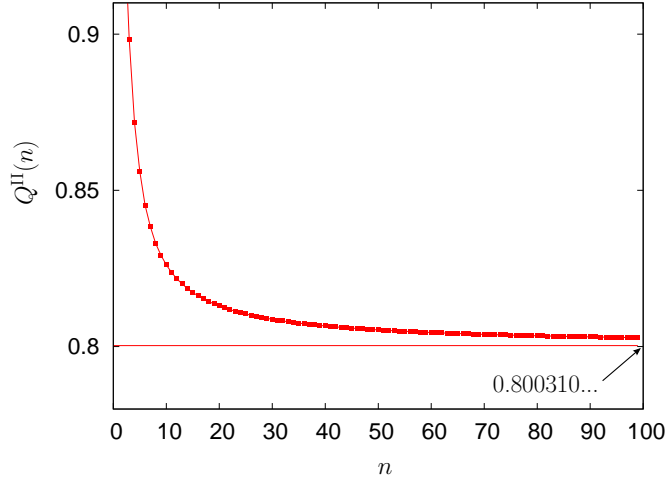}
\caption{The successive coefficients $Q^{\rm II}(n)$ of the series~(\ref{Q2_gf}) converge to 
$Q^{\rm II}(\infty)=0.800310\ldots$.
}
\label{fig_QII}
\end{figure} 

%
\bigskip{\bf Case III}. 
In this last case, a realization of the relevant set of ages is $\vec \ell = (\ell_1, \ell_2, \ldots, \ell_{m-1})$ and the joint probability $P^{\rm III}(\vec \ell, m,n)$ is thus given by integrating $P^{\rm I}(\zz_1,\ldots, \zz_{m-1}, a, m,n)$ on $a$, yielding
\begin{equation}\label{joint_pdf_III}
P^{\rm III}(\vec \ell, m,n) = f(\ell_1) f(\ell_2) \ldots f(\ell_{m-1})\, \sum_{a\ge0}q(a)\,
\delta\left(\sum_{k=1}^{m-1}\ell_k+a,n\right).
\end{equation}
For the cumulative distribution function of $\ell_{\max}^{\rm III}(n)$, $F^{\rm III}(\zz,n)$, we now have
\begin{eqnarray}
\sum_{n\ge0} F^{\rm III}(\zz,m,n) z^n = 
\Bigg(\sum_{\ll=1}^\zz f(\ll) z^\ll\Bigg)^{m-1}\sum_{a\ge0}q(a)z^{a}\;,
\end{eqnarray}
and therefore, summing on $m$,
\begin{eqnarray}\label{gf_F3}
\tilde F^{\rm III}(z,\zz)=\sum_{n\ge0}  F^{\rm III}(\zz,n) z^n=\frac{\tilde q(z)}{1 - \sum_{\ll=1}^\zz f(\ll) z^\ll} \;.
\end{eqnarray}
The normalization can be, once more, checked in these two last expressions by letting $\zz\to\infty$.
From (\ref{gf_F3}), we obtain the GF of $\langle \ell_{\max}^{\rm III}(n) \rangle = \sum_{\zz\ge0} (1 - F^{\rm III}(\zz,n))$ as
\begin{eqnarray}\label{lmax.3}
\sum_{n\ge0}  \langle \ell^{\rm III}_{{\rm \max}}(n) \rangle z^n 
&=& \sum_{\zz\ge0} \left[\frac{1}{1-z} - \frac{\tilde q(z)}{1 - \sum_{\ll=1}^\zz f(\ll)z^\ll} \right] \;,\\
&=& \frac{z^2}{2}+\frac{3z^3}{4}+z^4+\frac{5z^5}{4}+\frac{3z^6}{2}+\frac{895z^7}{512}+\cdots
\end{eqnarray}
As will be shown in the next section, $\langle\zz^{\rm III}_{\max}(n)\rangle$ scales as $n$ for large $n$.

As done for cases I and II, one can similarly obtain the GF of the probability of record breaking $Q^{\rm III}(n)$ as
\begin{eqnarray}\label{Q3_gf}
\tilde Q^{\rm III}(z)
&=&
\sum_{n\ge0} Q^{\rm III}(n) z^n = \tilde q(z) \sum_{\ll\ge1} \frac{f(\ll)z^\ll}{1 - \sum_{k=1}^\ll f(k)z^k} \;,\\
&=& \frac{z}{2}+\frac{5z^2}{8}+\frac{5z^3}{8}+\frac{77z^4}{128}+\cdots
\end{eqnarray}
In contrast with cases I and II, the coefficients of this series converge to zero, as will be analyzed in the next section (see figure~\ref{fig_numerics}).  

These explicit expressions of the GF in 
eqs.~(\ref{eq:lmaxI}, \ref{Q1_gf}, \ref{Q2_gf}, \ref{lmax.3}, \ref{Q3_gf}) are very convenient for an asymptotic analysis in the large $n$ limit. 
The identity~(\ref{eq:denominator}) will also be helpful. 
As we will see, although these formulas in 
eqs.~(\ref{eq:lmaxI}, \ref{lmax.3}) for the GF of $\langle \ell^\alpha_{\max}(n)\rangle$ and in eqs.~(\ref{Q1_gf}, \ref{Q2_gf}, \ref{Q3_gf}) for the GF of $Q^\alpha(n)$ look superficially similar in the different cases $\alpha = {\rm I, II}$ and III, a careful inspection shows that they actually give rise to quite different behaviors depending on $\alpha$. 

\section{Asymptotic analysis for a large number of steps $n$}\label{section:asymptotic}

To analyze the quantities $\langle \ell_{\max}^{\alpha}(n)\rangle$ and $Q^{\alpha}(n)$ in the large $n$ limit, we study the behavior of their associated GF in eqs.~(\ref{eq:lmaxI}, \ref{Q1_gf}, \ref{Q2_gf}, \ref{lmax.3}, \ref{Q3_gf}) in the limit $z \to 1$. 
To this purpose, it is convenient to set $z = \e^{-s}$ and study the limit $s \to 0$. 
In this limit, the discrete sums are replaced by integrals, as explained in \ref{appendixA}, and only the asymptotic behavior of $q(\ll)$ and $f(\ll)$ will matter in the following analysis:
\begin{eqnarray}\label{asympt}
q(\ll) \sim \frac{1}{\sqrt{\pi \ll}} \;, \; f(\ll) \sim \frac{1}{2\sqrt{\pi}\ll^{3/2}} \;.
\end{eqnarray}
Note that $q(j)$ decays algebraically for large $j$ as $q(j) \sim j^{-\theta}$ where $\theta = 1/2$ is the well known persistence exponent~\cite{Redner_book,Satya_review,Bray_review,bray94,derrida94} for the RW. 
The case of $Q^{\rm III}(n)$ deserves a separate treatment because all the structure of the series $\tilde q(z)$ and $\tilde f(z)$ matters for the evaluation of the subleading term, which is performed~in~\ref{subsec_app:III}.

\bigskip{\bf Case I}. 
In this case, the asymptotic analysis of the formulas for $\langle \ell_{\max}^{\rm I}(n)\rangle$ and $Q^{\rm I}(n)$ was carried out respectively in \cite{MZ2008, GMS2009}. It can be shown from eq.~(\ref{eq:lmaxI}), using the method explained in \ref{subsec_app:IandII}, that~\cite{MZ2008}
\begin{eqnarray}
\sum_{n\ge0}\e^{-s n} \langle \ell_{\rm max}^{I}(n)\rangle  = \frac{1}{s^2} C^{\rm I} + o (s^{-2})\;,
\end{eqnarray}
which implies $\langle \ell_{\rm max}^{I}(n) \rangle \sim C_{\rm I} \, n$ with $C_{\rm I} = Q^{\rm I}(\infty)$ as announced in eqs.~(\ref{Q_inf_I}, \ref{lmax_I}). Similarly, thanks to the relation (\ref{relation_I}), we find that $Q^{\rm I}(n) \to \langle \ell_{\max}^{\rm I}\rangle/n = Q^{\rm I}(\infty)$ \cite{GMS2009}, as given in eq.~(\ref{Q_inf_I}).

\bigskip{\bf Case II}. 
As mentioned above, the average value $\langle \ell_{\max}^{\rm II}(n)\rangle$ is infinite in this case and we thus focus here only on $Q^{\rm II}(n)$, whose GF is given in eq.~(\ref{Q2_gf}). By rewriting the denominator in eq.~(\ref{Q2_gf}) using the identity~(\ref{eq:denominator}) and the asymptotic behaviors in eq.~(\ref{asympt}), we obtain a relation in Laplace space (as shown in \ref{subsec_app:IandII})
\begin{eqnarray}
&&\sum_{n\ge0} \e^{-s n} Q^{\rm II}(n) = \frac{1}{s} Q^{\rm II}(\infty) + o(s^{-1})\;,
\label{laplace_qII} 
\end{eqnarray}
which implies the asymptotic behavior $\lim_{n \to \infty} Q^{\rm II}(n) = Q^{\rm II}(\infty)$
(see (\ref{eq:scheffer})).
This result, together with the value of $Q^{\rm II}(\infty)$ given by~(\ref{eq:scheffer}), 
was previously obtained by Scheffer \cite{Sch95} for Brownian motion, using rather complicated probabilistic methods. 
We have provided here a simple derivation of this constant. In addition, our approach shows that this result holds for random walks (\ref{def_RW}) with any jump distribution $p(\eta)$, including also L\'evy flights.

\bigskip{\bf Case III}. 
To compute the large $n$ behavior of $\langle \ell_{\max}^{\rm III}(n)\rangle$, we analyze the behavior of its GF in eq.~(\ref{lmax.3}) when $z = \e^{-s} \to 1$ (i.e. $s\to 0$). 
A simple analysis, along the lines explained in \ref{subsec_app:IandII} for $Q^\alpha(n)$, yields
\begin{eqnarray}\label{lmax.3_large}
&&\sum_{n\ge0} \e^{-sn} \langle \ell^{\rm III}_{{\rm \max}}(n) \rangle = \frac{1}{s^2} C^{\rm III} + o(s^{-2}) \;, 
\end{eqnarray}
hence $\langle \ell^{\rm III}_{{\rm \max}}(n) \rangle \sim C^{\rm III} n$, when $n \to \infty$, as announced in eq.~(\ref{C_III}). 

We then analyze the large $n$ behavior of the probability of record breaking $Q^{\rm III}(n)$. The analysis of its GF~(\ref{Q3_gf}) shows that, when $s\to0$ (see \ref{subsec_app:III} for details),
\begin{eqnarray}\label{eq:c0}
\sum_{n\ge0} \e^{-s\,n} Q^{\rm III}(n) \sim \frac{\ln{(1/\sqrt{s})}+c_0}{\sqrt{s}}  \;,
\end{eqnarray}
where $c_0=(\gamma+\ln(4/\pi))/2\approx0.409390\ldots$, which leads to
\begin{eqnarray}\label{Q3_asympt}
Q^{\rm III}(n) \underset {n \to \infty}{\sim}  \frac{\ln n+c}{2\sqrt{\pi n}}\;, \qquad c=2(\gamma+\ln 4)-\ln\pi,
\end{eqnarray}
where the amplitude ${1}/{(2 \sqrt{\pi})}$ and the constant $c\approx2.782290\ldots$ are universal ($\gamma$ is the Euler constant).

\begin{figure}[ht]
\centering
\includegraphics[width=\linewidth]{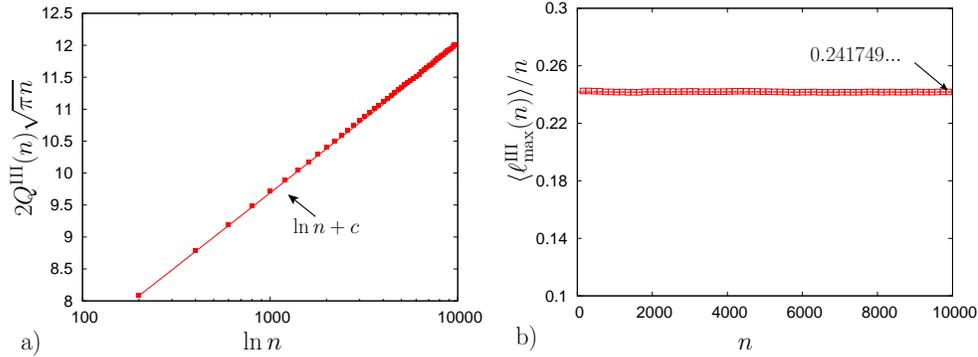}
\caption{
{\bf a)} Plot of $2 Q^{\rm III}(\n) \sqrt{\pi \n}$ as a function of $\ln \n$ for a RW with Gaussian jumps. 
The squares correspond to the results of numerical simulations while the solid line indicates the exact asymptotic result (\ref{Q3_asympt}).
{\bf b)} Plot of $\langle \ell_{\max}^{\rm III}(\n)\rangle/\n$ as a function of $\n$. 
The squares correspond to the results of the numerical simulation while the solid line is the exact value in 
eq.~(\ref{C_III}). }
\label{fig_numerics}
\end{figure}

In Fig.~\ref{fig_numerics}, we present the numerical results obtained for this third case concerning $Q^{\rm III}(n)$ (Fig. \ref{fig_numerics} a)) and $\langle \ell_{\max}^{\rm III}(n)\rangle/n$ (Fig. \ref{fig_numerics} b)). 
They show a very good agreement with our exact analytical results in eq.~(\ref{lmax.3_large}) and (\ref{Q3_asympt}). 
These numerical results have been obtained for a RW with a Gaussian jump distribution but we have checked that exactly identical results were obtained for different jump distributions $p(\eta)$, hence confirming the universality of these observables.

\section{Two other observables}\label{section:other_obs}

To complete our study, we study two other observables: the shortest age $\ell_{\rm min}^{\rm \alpha}(n) $  
defined as the smallest element of the sequences ${\cal A}^\alpha_{m,n}$ and the probability $Q_1^{\rm \alpha}(n)$ that the first record is the longest one. They are defined, respectively, as 
\begin{eqnarray}\label{def_lmin}
\ell_{\min}^{\rm \alpha}(\n) = 
\begin{cases}
& \min (\tau_1, \ldots, \tau_{\m-1}, A_n) \;, \; {\rm for} \; \alpha = {\rm I}\;, \\
&\min (\tau_1, \ldots, \tau_{\m-1}, \tau_\m) \;, \; {\rm for} \; \alpha = {\rm II}\;, \\
&\min (\tau_1, \ldots, \tau_{\m-1}) \;, \; {\rm for} \; \alpha = {\rm III} \;,
\end{cases}
\end{eqnarray} 
while 
\begin{eqnarray}
Q_1^{\rm \alpha}(\n) = 
\begin{cases}
&{\rm Prob} \, [\tau_1 \geq \max(\tau_2, \ldots, \tau_{\m-1},A_n)] \;, \; {\rm for \;} \alpha = {\rm I}\;,\\
&{\rm Prob} \, [\tau_1 \geq \max(\tau_2, \ldots, \tau_{\m})] \;, \; {\rm for \;} \alpha = {\rm II} \;,\\
&{\rm Prob} \, [\tau_{1} \geq \max(\tau_2, \ldots, \tau_{\m-1})] \;, \; {\rm for \;} \alpha = {\rm III} \;.\\
\end{cases}\label{def_Q1}
\end{eqnarray}
Below, we analyze separately these two quantities, both for finite $n$ and in the asymptotic limit $n \to \infty$. The main results are summarized in Table \ref{Table_2}.
\begin{table}
\begin{center}
\begin{tabular}{|c||c|c|}
\hline
Case & $\langle \ell_{\min}^{\alpha}(\n)\rangle$  & $Q_1^{\alpha}(\n)$ \\
\hline
$\quad$ & $\quad$ & $\quad$ \\
$\alpha$ = I & $\sim \sqrt{\dfrac{n}{\pi}}$ & $\dfrac{A^{\rm I}}{\sqrt{n}}$ \;, \; $A^{\rm I} = 0.962641\ldots$ \\
$\quad$ & $\quad$ & $\quad$ \\
\hline
$\quad$ & $\quad$ & $\quad$ \\
$\alpha$ = II & $\infty$ & $\dfrac{A^{\rm II}}{\sqrt{n}}$ \;, \; $A^{\rm II} = 0.772063\ldots$ \\
$\quad$ & $\quad$ & $\quad$ \\
\hline
$\quad$ & $\quad$ & $\quad$ \\
$\alpha$ = III & $\sim \dfrac{3}{2}$  & $\sim  \dfrac{\ln \n+c}{2\sqrt{\pi\n}}$ \\
 $\quad$ & $\quad$ & $\quad$ \\
\hline
\end{tabular}
\caption{Asymptotic behaviors, when $\n \to \infty$, of $\langle \ell_{\min}^\alpha(n)\rangle$ and $Q^{\alpha}_1(n)$ studied in section \ref{section:other_obs} in each three cases (\ref{def_I}, \ref{def_II}, \ref{def_III}).
The constants $A^{\rm I}$ and $A^{\rm II}$ are given respectively in~(\ref{expr_AI}) and~(\ref{def_AII}) while $c$ is defined in~(\ref{Q3_asympt}).}
\label{Table_2}
\end{center}
\end{table}

\subsection{The shortest age $\ell_{\min}^{\alpha}(n) $}

The common scheme for the calculation of the distribution of $\ell_{\min}^{\alpha}(n)$ is as follows.
Its complementary cumulative distribution function is defined by 
\begin{eqnarray}
\fl G^{\alpha}(\zz,m,n)=\prob(\ell_{\min}^{\alpha}(n)\ge\zz,R_n=m),\\
\fl G^{\alpha}(\zz,n)=\prob(\ell_{\min}^{\alpha}(n)\ge\zz)=\sum_{m\ge1}\prob(\ell_{\min}^{\alpha}(n)\ge\zz,R_n=m)\;.
\end{eqnarray}
From $G^{\alpha}(\zz,n)$, we obtain the average value of $\ell_{\rm min}^{\rm \alpha}(n)$ as
\begin{equation}
\langle\ell_{\min}^{\alpha}(n)\rangle=\sum_{\zz\ge\textcolor{black}{1}}G^{\alpha}(\zz,n) \;.
\end{equation}
Its GF with respect to $n$ reads
\begin{equation}\label{eq:fg_lmin}
\sum_{n\ge0}\langle\ell_{\min}^{\alpha}(n)\rangle z^n=\sum_{\zz\ge1}\tilde G^{\alpha}(z,\zz) \;, \; \tilde G^{\alpha}(z,\zz) = \sum_{\zz\ge\textcolor{black}{1}}G^{\alpha}(\zz,n)z^\ell \;,
\end{equation}
whose analysis depends on the different cases $\alpha = {\rm I, II}$ and $\alpha= {\rm III}$.  

\bigskip{\bf Case I}. 
In this case 
$\ell_{\min}^{\rm I}(n)$ as defined in eq.~(\ref{def_lmin}) takes values between $0$ and $n$, because 
if there is a record at the last step, then $\ell_{\min}^{\rm I}=A_n=0$,
and if there are no record beyond the first one, i.e. $m=1$, then 
$\ell_{\min}^{\rm I}=A_n$ which is equal to $n$.
We thus obtain
\begin{eqnarray}\label{eq:gfImin}
\sum_{n\ge0}G^{\rm I}(\zz,m,n)z^n=\sum_{j\ge\zz}q(j)z^j \left(\sum_{j\ge\zz}f(j)z^j\right)^{m-1},
\end{eqnarray}
from which it follows that
\begin{eqnarray}\label{eq:GminI}
\tilde G^{\rm I}(z,\zz)=\frac{\sum_{j\ge\zz}q(j)z^j}{1-\sum_{j\ge\zz}f(j)z^j}.
\end{eqnarray}
The normalization of the distribution of $\ell_{\min}^{\rm I}(n)$ can be checked by setting $\zz=0$ in this expression.
We thus get
\begin{equation}
\sum_{n\ge0}\langle\ell_{\min}^{I}(n)\rangle z^n=
\frac{z}{2}+z^2+\frac{21z^3}{16}+\frac{51z^4}{32}+\frac{461z^5}{256}+\cdots
\end{equation}

The asymptotic analysis of~(\ref{eq:GminI}), when $z\to1$, gives, setting $z = \e^{-s}$
\begin{equation}
\sum_{n\ge0}\langle\ell_{\min}^{\rm I}(n)\rangle \e^{-s n}\sim\frac{1}{2s^{3/2}},
\end{equation}
hence we recover the result of \cite{MZ2008}
\begin{equation}\label{eq:lminI}
\langle\ell_{\min}^{\rm I}(n)\rangle\sim\sqrt{\frac{n}{\pi}}.
\end{equation}
In this case, the average value $\langle \ell^I_{\min}\rangle$ is dominated by the paths with a single record, $m=1$, occurring at $x_0 = 0$. 
This can be seen on~(\ref{eq:gfImin}) where the term with $m=1$ gives the 
dominating contribution to~(\ref{eq:GminI}).
The result~(\ref{eq:lminI}) can then be simply recovered by noting that a path with $m=1$ is such that it stays negative
up to step $n$. Such paths occur with a probability $q(n)\sim 1/\sqrt{\pi n}$ and they contribute to a value of $\ell_{\min} = n$, implying the result in (\ref{eq:lminI}).
This shows explicitly that $\langle \ell^{\rm I}_{\min}(n)\rangle$ is
dominated by rare events, such that the random walk never crosses the origin up to step $n$.

\bigskip{\bf Case II}. 
In this case, we have from eq.~(\ref{eq:fg_lmin})
\begin{eqnarray}\label{eq:GminII}
\tilde G^{\rm II}(z,\zz)=
\frac{1}{1-z}\frac{\sum_{j\ge\zz}f(j)(1-z^j)}{1-\sum_{j\ge\zz}f(j)} \;.
\end{eqnarray}
The normalization of the distribution of $\ell_{\min}^{\rm II}(n)$ can be checked by setting $\zz=0$.
The event $m=1$ is still dominating, which implies that $\ell_{\min}^{\rm II}=\tau_1$, and hence its first moment $\langle \ell_{\min}^{\rm II} \rangle$
is not defined. 

\bigskip{\bf Case III}. 
In this third case, if $m=1$, the sequence is empty.
We take conventionally $\ell_{\min}^{\rm III}(n)=0$ in such an instance.
Following the same method as above, we obtain
\begin{eqnarray}\label{eq:GminIII}
\tilde G^{\rm III}(z,\zz)=
\tilde q(z)\left(\delta(\ell,0)+\frac{\sum_{j\ge\zz}f(j)z^j}{1-\sum_{j\ge\zz}f(j)z^j}\right) \;.
\end{eqnarray}
The normalization of the distribution of $\ell_{\min}^{\rm III}(n)$ can be checked by setting $\zz=0$ in this expression.
From~(\ref{eq:GminIII}) it follows that
\begin{eqnarray}\label{eq:lminIII}
\sum_{n\ge0}\langle\ell_{\min}^{\rm III}(n)\rangle z^n
&=&\sum_{\zz\ge1}\tilde G^{\rm III}(z,\zz),\\
&=&\frac{z}{2}+\frac{3z^2}{4}+\frac{7z^3}{8}+\frac{31z^4}{32}+\frac{33z^5}{32}+\cdots,
\end{eqnarray}
which we have to analyze in the limit $z \to 1$. A close inspection of the probability distribution 
of $\ell_{\min}^{\rm III}(n)$ shows that it is made of two contributions.
The first one comes from $\ell_{\min}^{\rm III}(n)=1$, the other one from all values of $\ell_{\min}^{\rm III}(n)$ larger than 1. One can show that for large $n$ this distribution admits the following expansion:
\begin{eqnarray}\label{dist_lmin}
{\rm Prob}(\ell_{\min}^{\rm III}(n) = j) = \delta(j,1) + \frac{1}{\sqrt{n}}\varphi(j) + {\cal O}(n^{-1}) \;,
\end{eqnarray}
where $\varphi(j) \propto j^{-3/2}$ for large $j$. Because of this slow algebraic decay, one has that $\sum_{j=1}^n j \varphi(j) \sim \sqrt{n}$ and hence the two first terms in this expansion (\ref{dist_lmin}) --and maybe the higher order terms which we have not tried to evaluate-- lead to a constant contribution to $\langle\zz_{\min}^{\rm III}(n)\rangle$ that we now evaluate starting directly from (\ref{eq:GminIII}).

Guided by the above observation (\ref{dist_lmin}) we treat separately the term $\ell = 1$ in the sum in (\ref{eq:lminIII}):
\begin{eqnarray}\label{lmin_decomp}
\sum_{n\ge0}\langle\ell_{\min}^{\rm III}(n)\rangle z^n = \tilde G^{\rm III}(z,1) + \sum_{\ell \geq 2} G^{\rm III}(z,\ell) \;.
\end{eqnarray}
The first term $\tilde G^{\rm III}(z,1)$ is easily analyzed when $z \to 1$, , setting $z = \e^{-s}$ 
%
\begin{equation}\label{lmin_part1}
\tilde G^{\rm III}(z,1)=\frac{1-\sqrt{1-z}}{1-z}\sim \frac{1}{s}\;,\; s \to 0 \;.
\end{equation}
Hence for $n$ large
\begin{equation}
G^{\rm III}(1,n)=\prob(\zz_{\min}^{\rm III}(n)\ge1)\to1.
\end{equation}
On the other hand, the sum over $\ell \geq 2$ in eq.~(\ref{lmin_decomp}) can be approximated by an integral in the limit $z = \e^{-s} \to 1$, as done for the analysis of $Q^\alpha(n)$ along the lines explained in \ref{subsec_app:IandII}, yielding
\begin{eqnarray}\label{lmin_part2}
\sum_{\ell \geq 2} G^{\rm III}(z,\ell) \sim \frac{1}{s} \frac{1}{2\sqrt{\pi}}\int_0^\infty {\rm d}y 
\int_y^\infty{\rm d}x\, \frac{\e^{-x}}{x^{3/2}}   = \frac{1}{2s} \;, \; s \to 0 \;.
\end{eqnarray} 
Hence summing up these two contributions (\ref{lmin_part1}) and (\ref{lmin_part2}) we obtain from (\ref{lmin_decomp}) that
\begin{eqnarray}\label{eq:lmin_III}
\lim_{n \to \infty} \langle \ell^{\rm III}_{\min}(n) \rangle = 1 + \frac{1}{2} = \frac{3}{2} \;,
\end{eqnarray}
as given Table \ref{Table_2}. Again, by comparing the behavior of $\langle \ell^{\rm I}_{\min}(n) \rangle \sim \sqrt{n/\pi}$ in (\ref{eq:lminI}) and $\langle \ell^{\rm III}_{\min}(n) \rangle \sim 3/2$ in (\ref{eq:lmin_III}) we see that modifying the last element of the sequences of the ages of the records has drastic consequences on its minimal element.


\subsection{Probability that the longest lasting record is the first one}

As we have done before for the other observables, we introduce the quantity $Q^{\rm \alpha}_1(m,n)$, which is the joint probability that the first age is the longest of the sequences (as in eq.~(\ref{def_Q1})) and that there are $R_n = m$ records. The total probability $Q_1^{\rm \alpha}(n)$ is simply computed by summation over $m$ as  
\begin{eqnarray}
Q^{\rm \alpha}_1(n) = \sum_{m=1}^n Q^{\rm \alpha}_1(m,n) \;.
\end{eqnarray} 

{\bf Case I}. In this case (\ref{joint_caseI}), the configuration with one single record, $m=1$, in $x_0 = 0$ has to be treated separately. In such configuration the first age, namely $A_n$, is indeed the longest one. This yields
\begin{eqnarray}\label{eq:Q1_m1}
Q_1^{\rm I}(m=1,n) = q(n) \;.
\end{eqnarray}
For $m\geq 2$, $Q^{\rm I}_1(n)$ is given by summing up $P^{\rm I}(\vec \ell,m,n)$ (\ref{joint_caseI}) over the relevant configurations of ages $\vec \ell$ as
\begin{eqnarray}\label{eq:Q1_m}
Q^{\rm I}_1(m,n) = \sum_{\ell_1\geq 1} \sum_{\ell_2 = 1}^{\ell_1}  \cdots \sum_{\ell_{m-1} = 1}^{\ell_1} \sum_{a=0}^{\ell_1} P^{\rm I}(\vec \ell,m,n) \;, \; m \geq 2 \;.
\end{eqnarray}
From eqs.~(\ref{eq:Q1_m1}, \ref{eq:Q1_m}), we obtain the GF $\tilde Q_1^{\rm I}(z)$ as
\begin{equation}\label{eq:gf_q11}
\tilde Q_1^{\rm I}(z) = \sum_{n \geq 1}Q_1^{\rm I}(n)z^n = \frac{1}{\sqrt{1-z}} + \sum_{j \geq 1} \frac{f(j) z^j}{1 - \sum_{k=1}^j f(j)z^j}\sum^j_{a=0} q(a)z^a \;,
\end{equation} 
from which the first terms $Q_1^{\rm I}(1), Q_1^{\rm I}(2) \cdots$ can be read off directly
\begin{eqnarray}
\tilde Q_1^{\rm I}(z) = 1 + z + z^2 + \frac{3}{4}z^3 + \frac{5}{8}z^4 + \cdots \;.
\end{eqnarray}
To obtain the large $n$ behavior of $Q_1^{\rm I}(n)$, we analyze the GF $\tilde Q_1^{\rm I}(z)$ in the limit $z \to 1$, where we set $z = \e^{-s}$, as we have done before in section \ref{section:asymptotic}. In this limit, the discrete sums in (\ref{eq:gf_q11}) can be replaced by integrals, as explained in \ref{subsec_app:IandII} for the analysis of $Q^\alpha(n)$, which finally yields
\begin{eqnarray}
&&\sum_{n \geq 0} Q_1^{\rm I}(n) \e^{- s n} \sim \sqrt{\frac{\pi}{s}} A^{\rm I} \;, \; s \to 0 \;, \\
&&A^{\rm I} = \frac{1}{\sqrt{\pi}} \left(1 + \frac{1}{2} \int_0^\infty \frac{{\rm d}x}{x} \frac{{\rm erf}(\sqrt{x})}{1+ \sqrt{\pi x} \, \e^x \, {\rm erf}{(\sqrt{x})}} \right) = 0.962641\ldots \;, \label{expr_AI}
\end{eqnarray}
implying $Q_1^{\rm I}(n) \sim A^{\rm I}/\sqrt{n}$, as announced in Table \ref{Table_2}. 

\bigskip
{\bf Case II}. In this case also (\ref{eq:joint_pdf_II}, \ref{def_Q1}), the configuration with one single record, $m=1$, in $x_0 = 0$ has to be treated separately. In this case, the age $\ell_1 > n$ is obviously the largest one and therefore
\begin{eqnarray}\label{eq:Q2_m1}
Q^{\rm II}_1(m=1,n) = \sum_{j > n} f(j) = q(n) \;,
\end{eqnarray}
where we have used $f(j) = q(j-1) - q(j)$. For $m\geq 2$, $Q_1^{\rm II}(m,n)$ is given by summing up $P^{\rm II}(\vec \ell,m,n)$ (\ref{eq:joint_pdf_II}) over the relevant configurations of ages $\vec \ell$ as
\begin{eqnarray}\label{eq:Q2_m}
Q_1^{\rm II}(m,n) = \sum_{\ell_1\geq 1} \sum_{\ell_2 = 1}^{\ell_1} \cdots \sum_{\ell_m = 1}^{\ell_1} P^{\rm II}(\vec \ell,m,n) \;.
\end{eqnarray}
From eqs.~(\ref{eq:Q2_m1}, \ref{eq:Q2_m}) we obtain the GF $\tilde Q_1^{\rm II}(z)$ as
\begin{equation}\label{q1_2gf}
\tilde Q_1^{\rm II}(z) = \frac{1}{\sqrt{1-z}}+\sum_{n \geq 0}Q_1^{\rm II}(n) z^n = \sum_{j \geq 1} \frac{f(j)z^j}{1 - \sum_{k=1}^{j} f(k) z^k} \sum_{\ell=1}^j f(\ell) \frac{1-z^\ell}{1-z} \;,
\end{equation}
from which the first terms $Q_1^{\rm II}(1), Q_1^{\rm II}(2), \ldots$ can be read off directly
\begin{eqnarray}
\tilde Q_1^{\rm II}(z) = 1 + \frac{3}{4}z + \frac{37}{64}z^2 + \frac{121}{256}z^3 + \frac{6609}{16384} z^4 + \ldots 
\end{eqnarray}
On the other hand, when $z \to 1$, setting $z = \e^{-s}$, the expression in (\ref{q1_2gf}) can be analyzed along the lines detailed in \ref{subsec_app:IandII} to yield
\begin{eqnarray}
&&\sum_{n\geq0} Q_1^{\rm II}(n) \e^{-s n } \sim \sqrt{\frac{\pi}{s}} A^{\rm II} \;, \, s \to 0 \;, \\
&&A^{\rm II} = \frac{1}{\sqrt{\pi}} \left(1 + \frac{1}{2 \sqrt{\pi}}\int_0^\infty \frac{{\rm d}x}{x^{3/2}} \frac{\sqrt{\pi x}\, {\rm erf}(\sqrt{x}) - (1-\e^{-x}) }{1 + \sqrt{\pi x} \e^x \, {\rm erf}(\sqrt{x})}   \right) \label{def_AII}\\
&& \hspace*{0.5cm}= 0.772063 \ldots \;.
\end{eqnarray}
implying $Q_1^{\rm II}(n) \sim A^{\rm II}/\sqrt{n}$, as announced in Table \ref{Table_2}.

\bigskip
{\bf Case III}. In this case, all the ages, $\tau_1, \cdots, \tau_{m-1}$, of the sequence ${\cal A}^{\rm III}_{m,n}$ defined in~(\ref{def_III}) are statistically equivalent. Therefore the probability that the first one, $\tau_1$, or the last one, $\tau_{m-1}$, (or any other $\tau_k$) is the longest one is actually the same, independently of $k$. Hence one has 
\begin{eqnarray}
Q_1^{\rm III}(n) = Q^{\rm III}(n) \;,
\end{eqnarray} 
which, together with the asymptotic results for $Q^{\rm III}(n)$ given in Table \ref{Table_1}, yields immediately the results for $Q_1^{\rm III}(n)$ given in Table \ref{Table_2}.  

%

%

\section{Conclusion and perspectives}

To conclude, we have proposed and studied three distinct sequences $\alpha = $ I, II and III, of the ages of the records of a RW up to step $\n$ (\ref{def_I}, \ref{def_II}, \ref{def_III}). In each case, the various elements of these sequences, i.e. the ages, are independent random variables {\it except} for a global constraint (the Kronecker delta in eqs.~(\ref{joint_caseI}, \ref{joint_pdf_III}) and the indicator function in eq.~(\ref{eq:joint_pdf_II})). This global constraint induces correlations which, as we have shown here, are responsible for a statistical behavior of the ages of the records of a RW which differs significantly from the one of independent random variables. A second important feature is that these sequences of the ages of the records are extremely sensitive to the last record. The mechanism behind this high sensitivity is that many observables associated to the ages are dominated by rare events, whose statistics is controlled, to a large extent, precisely by the last record duration. Finally the statistics of the ages of the records is completely universal, i.e. independent of the jump distribution $p(\eta)$, including L\'evy flights, provided it is continuous. We emphasize that the origin of universality found here is completely different from the case of records for i.i.d. random variables (where it stems from its connection with the statistics of permutations): here universality has its roots in the celebrated Sparre Andersen theorem \cite{SA53,SA54}. The case of discrete RW needs to be treated separately \cite{MZ2008} as in this case the expression for $q(n)$ and $f(n)$ gets modified \cite{MZ2008,Bray_review}: one still has $q(n) \propto n^{-1/2}$ and $f(n) \propto n^{-3/2}$ for large $n$ but with different prefactors. 
Therefore the behavior with $n$ of the different observables computed here are the same for discrete RW but (for some observables \cite{MZ2008}) with different amplitudes. 
We emphasize that 
all the formulas which we have derived for the GF of the observables studied here in terms of the GF $\tilde q(z)$ and $\tilde f(z)$ remain valid in the discrete case. 
Hence the study of discrete RW is a simple extension of the computations presented in this paper.   

We have illustrated these remarkable features of the sequence of ages of records by focusing on two important quantities: (i) the probability $Q^{\rm \alpha}(\n)$ of record breaking of the longest age (of a record) at step $\n$, and the age of the longest lasting record $\ell_{\max}^\alpha(\n)$. 
Importantly, the large $\n$ behaviors of these quantities give rise to three universal constants (see Table \ref{Table_1}). 
While the two first ones had appeared before in the excursion theory of Brownian motions \cite{Sch95,PY97} --for which we provide here a much simpler derivation-- the third constant $C^{\rm III} = 0.241749 \ldots$ has not been discussed before in the literature. 
We have also discussed two additional observables, the minimal age of the sequence $\ell^\alpha_{\min}$ and the probability $Q_1^{\alpha}(n)$ that the first age is the longest one. They also show a universal behavior which is again quite sensitive to the last record. We have shown that 
they give rise to additional universal constants (see Table \ref{Table_2}).

Besides the case of discrete RW, there are several natural extensions of the present work. 
First, it would be interesting to study the effects of a constant drift on these quantities $Q^{\rm \alpha}(\n)$ and 
$ \ell^{\alpha}_{\max}(\n)$ (as well as on $Q^{\rm \alpha}_1(\n)$ and $ \ell^{\alpha}_{\min}(\n)$). 
For case I, this study was carried out in Ref.~\cite{MSW2012} where it was shown that the presence of a finite drift $|c| > 0$ induces five distinct universality classes (at variance with a single one when $c=0$) depending on the sign of the drift and the L\'evy index $\mu$ of the jump distribution $p(\eta)$, leading to a rich behavior of these quantities in the $(c, \mu)$ plane \cite{MSW2012}. 
For cases II and III, one expects naturally that the behaviors of $Q^{\alpha}(\n)$ and $\ell_{\max}^{\alpha}(\n)$ will be different in these five distinct regions of the $(c, \mu)$ plane, which remain to be studied in detail. 
It will also be interesting to extend the study of cases II and III to other stochastic processes, including in particular non-Markovian ones like the randomly accelerated process or the fractional Brownian motion, for which the study of case I had revealed interesting features \cite{GMS2009,GRS2010}. 

Finally, it is worth mentioning that the formalism used in the present work in order to compute the distributions of 
$\ell_{\max}^{\alpha}(n)$ and $\ell_{\rm min}^{\alpha}(n)$, or the probabilities $Q^{\alpha}(n)$, $Q_{1}^{\alpha}(n)$, relies on the renewal properties of the sequences of ages ${\cal A}^{\alpha}_{m,n}$.
A natural extension of the study performed here consists in considering the statistics of the same quantities for the sequences of the excursions of stochastic processes given by a renewal process, like, e.g. Brownian motion.
This study was already addressed in~\cite{GMS2009}.
A more systematic account of this theory will be given elsewhere~\cite{ustocome}.
There, the emphasis will be on the role of the {\it distribution of excursions $\tau_1,\tau_2,\ldots$} on the behaviour of the quantities studied in the present work, and in particular on the universal or non universal properties of these quantities depending on the nature of the distribution of excursions.
 
\ack
 SNM and GS acknowledge support by ANR grant
2011-BS04-013-01 WALKMAT and in part by the Indo-French Centre for the Promotion of Advanced Research under Project 4604-3. GS acknowledges support from Labex-PALM (Project Randmat). 

\appendix

\section{Asymptotic analysis of $Q^{\alpha}(n)$}
\label{appendixA}

In this appendix, we show how to obtain the large $n$ behavior of $Q^{\alpha}(n)$. 
We analyze separately cases I and II --which can both be analyzed exactly in the same way-- and case III which necessitates a separate (more detailed) analysis to extract the subleading corrections and the constant $c$ in eq.~(\ref{Q3_asympt}). The analysis presented for $\alpha = {\rm I, II}$ can be easily adapted to study the asymptotic behavior of the GF of the other observables studied in this paper.

\subsection{Cases I and II}
\label{subsec_app:IandII}

Both cases can be treated along the same lines, which we illustrate on case ${\rm II}$, as this yields a new and simpler derivation of Scheffer's result \cite{Sch95} (while a similar analysis of the case I, for continuous time processes, was carried out in \cite{GMS2009}). 
To obtain the large $n$ behavior of $Q^{\rm II}(n)$, we study the behavior of its GF with respect to $n$, $\tilde Q^{\rm II}(z)$ given in eq.~(\ref{Q2_gf}), in the limit $z \to 1$. 
Setting $z = \e^{-s}$ we thus have from (\ref{Q2_gf})  
\begin{eqnarray}\label{Q2_gf_app}
\sum_{n\ge0} Q^{\rm II}(n) \e^{-s n} = \frac{1}{1-\e^{-s}} \sum_{j\ge1} \frac{f(j)(1 - \e^{-s j})}{1 - \sum_{k=1}^j f(k) \e^{-s k}} \;,
\end{eqnarray} 
which we study in the limit $s \to 0$. To compute the small $s$ expansion of (\ref{Q2_gf_app}) it is useful to rewrite the denominator using the identity in eq.~(\ref{eq:denominator}): 
\begin{eqnarray}\label{eq:denominator_app}
1 - \sum_{k=1}^j f(j)z^j = q(j) z^j+ (1-z)\sum_{k=0}^{j-1}q(j)z^j  \;,
\end{eqnarray}
such that 
\begin{eqnarray}\label{Q2_gf_app_step2}
\sum_{n\ge0} Q^{\rm II}(n) \e^{-s n} \sim \frac{1}{s} \sum_{j\ge1} \frac{f(j)(1 - \e^{-s j})}{s \sum_{k=0}^{j-1} q(k)\e^{-sk} + q(j)\e^{-s j}  } \;,
\end{eqnarray} 
where we have simply expanded $1 - \e^{-s} = s + {\cal O}(s^2)$. The next step is to realize that when $s \to 0$ the discrete sums in (\ref{Q2_gf_app_step2}) can be replaced, at leading order, by integrals. Hence, for instance, in the denominator one has:
\begin{eqnarray}\label{denominator_EML}
s \sum_{k=0}^{j-1} q(k)\e^{-sk} \sim s \int_0^{j-1}{\rm d}k\, q(k) \e^{-s k}  \;,
\end{eqnarray}
where we have retained only the first term of the Euler-Mac Laurin expansion of the discrete sum (one can show that higher order terms lead to subleading contributions when $s \to 0$). Performing the change of variable $y = s k$ in (\ref{denominator_EML}) and using that $q(y/s) \sim \sqrt{s/(\pi y)}$ as $s \to 0$ (\ref{asympt}), one arrives finally at
\begin{eqnarray}\label{denominator_final}
s \sum_{k=0}^{j-1} q(k)\e^{-sk} \sim \sqrt{\frac{s}{\pi}} \int_0^{s \, j}{\rm d}y\, y^{-1/2} \e^{-y}  = \sqrt{s} \; {\rm erf}(\sqrt{s j})\;,
\end{eqnarray} 
in the limit $s \to 0$, $j \to \infty$, keeping the product $s \, j$ fixed. We can then inject this expression (\ref{denominator_final}) into eq.~(\ref{Q2_gf_app_step2}) and replace the discrete sum over $j$ in (\ref{Q2_gf_app_step2}) by an integral, just as we did above in (\ref{denominator_EML}):
\begin{eqnarray}\label{Q2_gf_app_step2bis}
\sum_{n\ge0} Q^{\rm II}(n) \e^{-s n} = \frac{1}{s} \int_1^\infty{\rm d}j  \frac{f(j)(1 - \e^{-s j})}{\sqrt{s} \, {\rm erf}(\sqrt{s j}) + q(j) \e^{-s j}} \;.
\end{eqnarray} 
Finally, performing the change of variable $j = y/s$ and using that $f(y/s) \sim (s/y)^{3/2}/(2 \sqrt{\pi})$ as $s\to 0$, one finally arrives at
\begin{eqnarray}
&&\sum_{n\ge0} \e^{-s n} Q^{\rm II}(n) \sim \frac{1}{s} Q^{\rm II}(\infty) \;, \nonumber \\
&&Q^{\rm II}(\infty) = \frac{1}{2} \int_0^\infty {\rm d}x\, \frac{\e^x-1}{x + \sqrt{\pi }x^{3/2} \, \e^x \, {\rm erf}(\sqrt{x})} = 0.800310 \ldots \;,
\end{eqnarray}
as announced in the text (\ref{eq:scheffer}, \ref{laplace_qII}).

\subsection{Case III}
\label{subsec_app:III}

To obtain the large $n$ behavior of $Q^{\rm III}(n)$ for large $n$, we have to analyze the GF $\tilde Q^{\rm III}(z)$ given in eq.~(\ref{Q3_gf}) in the limit $z \to 1$. 
We recall that
\begin{eqnarray}\label{Q3_gf_app}
\tilde Q^{\rm III}(z)
= \tilde q(z) \sum_{\ll\ge1} \frac{f(\ll)z^\ll}{1 - \sum_{k=1}^\ll f(k)z^k}
=\tilde q(z) S(z)
 \;,
\end{eqnarray}
where we have denoted the discrete sum in (\ref{Q3_gf_app}) by $S(z)$.
In this case, we not only want to get the leading term for large $n$, which is $\propto \ln{n}/\sqrt{n}$ (which can be easily obtained along the lines explained above in \ref{subsec_app:IandII})  but also the first correction, $\propto 1/\sqrt{n}$, which gives rise to a nontrivial amplitude $c$ which we show how to compute here. 

We first rewrite the denominator of $S(z)$ as
\begin{equation}
1- \sum_{k=1}^\ll f(k) z^k = 1 - \tilde f(z)+ \sum_{k\geq \ll +1} f(k)z^k = \sqrt{1-z} +  \sum_{k\geq j +1} f(k)z^k
\end{equation}
such that
\begin{eqnarray}\label{Q3_gf_app_step2}
S(z) =\sum_{j\geq 1} \frac{f(j) z^j}{\sqrt{1-z} + \sum_{k\geq \ll +1} f(k)z^k} \;.
\end{eqnarray}
The limiting value of this denominator, when $z\to1$, follows immediately from the identity (\ref{eq:denominator})
\begin{eqnarray}
\lim_{z \to 1} \left(\sqrt{1-z} + \sum_{k\geq \ll +1} f(k)z^k\right) = q(j) \;.
\end{eqnarray}
We therefore introduce the function $J(z)$ such that
\begin{equation}
J(z) = \sum_{j \geq 1} \frac{f(j)}{q(j)}z^j 
=  \sum_{j \geq 1} \frac{z^j}{2j-1} 
= \frac{1}{2}\sqrt{z}\, \ln \frac{(1+\sqrt{z})^2}{1-z}
\;,
\end{equation}
where we have used $f(j)/q(j) = 1/(2j-1)$. 
This function bears the logarithmic divergence of $S(z)$:
\begin{equation}\label{eq:jz}
J(z)\sim\ln 2+\frac{1}{2}\ln \frac{1}{1-z},\quad (z\to1).
\end{equation}
Now, the difference
\begin{equation}\label{diff}
S(z)- J(z) = \sum_{j \geq 1} f(j) z^j \left[ \frac{1}{\sqrt{1-z}+ \sum_{k\geq \ll +1} f(k)z^k} - \frac{1}{q(j)}\right] \;,
\end{equation}
can be safely approximated, when $z\to1$, by an integral (this is guaranteed by the fact that the summand in (\ref{diff}) is identically zero for all values of $j$ when $z \to 1$). 
Performing the same kind of manipulations as in \ref{subsec_app:IandII} we show that as $z=\e^{-s}\to1$ (or $s \to 0$)
\begin{equation}
\fl S(z) - J(z) \sim c_1 \;, \quad c_1 = \frac{1}{2} \int_0^\infty \frac{{\rm d}x}{x} \e^{-x}\left(\frac{1}{\e^{-x} + \sqrt{\pi x} \, {\rm erf}(\sqrt{x})} - 1\right) \;.
\end{equation} 
This last integral can eventually be evaluated in a closed form (noticing that 
$\frac{d}{dx}[\e^{-x} + \sqrt{\pi x}\,{\rm erf}(\sqrt{x})] = \sqrt{\pi/(4x)}\, {\rm erf}(\sqrt{x})$):
\begin{eqnarray}
c_1 = \frac{1}{2} \left(\gamma - \ln \pi \right) \;,
\end{eqnarray}
where $\gamma$ is the Euler constant. 

Finally, using~(\ref{eq:jz}), one obtains that
\begin{equation}
S(z)\sim c_1+\ln2+\frac{1}{2}\ln \frac{1}{1-z},\quad(z\to1),
\end{equation}
and therefore, 
\begin{eqnarray}\label{eq:c0_app}
\sum_{n\ge0} \e^{-s\,n} Q^{\rm III}(n) \sim \frac{\ln{(1/\sqrt{s})}+c_0}{\sqrt{s}}  \;,
\end{eqnarray}
where $c_0=(\gamma+\ln(4/\pi))/2\approx0.409390\ldots$, as announced in the text below eq.~(\ref{eq:c0}). By inverting the Laplace transform (\ref{eq:c0_app}), we obtain the result given in eq.~(\ref{Q3_asympt}).

\section*{References}

\end{document}